\documentclass[preprint]{raa}            

\usepackage{graphicx,times}   
\usepackage{natbib}
\usepackage{amssymb,amsmath}
\bibpunct{(}{)}{;}{a}{}{,}

\begin{document}

\title{Effect of non-stationary accretion on spectral state transitions: An example of a persistent neutron star LMXB 4U 1636-536
}

   \volnopage{Vol.0 (200x) No.0, 000--000}      
   \setcounter{page}{1}          

   \author{H. Zhang
      \inst{1}
   \and W.-F. Yu
      \inst{1}
   }

   \institute{Key Laboratory for Research in Galaxies and
    Cosmology, Shanghai Astronomical Observatory, Chinese Academy of
    Sciences, 80 Nandan Road, Shanghai 200030, China; {\it hzhang@shao.ac.cn, zhanghuienator@gmail.com}\\
     }

   \date{Received~~2017 month day; accepted~~2017~~month day}

\abstract{Observations of the black hole and neutron star X-ray binaries show that the luminosity of the hard-to-soft state transition is usually higher than that of the soft-to-hard state transition, indicating additional parameters other than the mass accretion rate is required to interpret spectral state transitions. It has been found in some individual black hole or neutron star soft X-ray transients that the luminosity corresponding to the hard-to-soft state transition is positively correlated with the peak luminosity of the following soft state. In this work, we report the discovery of the same correlation in a single persistent neutron star low mass X-ray binary (LMXB) 4U1636-536 based on the data from the All Sky Monitor(ASM) on board the {\it RXTE}, the Gas Slit Camera (GSC) on board the {\it MAXI} and the Burst Alert Telescope (BAT) on board the {\it Swift}.  We also found such a positive correlation holds in the persistent neutron star LMXB in a luminosity range spanning by about a factor of four. Our results indicates that non-stationary accretion also plays an important role in driving X-ray spectral state transitions in persistent accreting systems with small accretion flares, which is much less dramatic compared with the bright outbursts seen in many Galactic LMXB transients.
\keywords{accretion, accretion disks --- X-rays: binaries --- stars: neutron}
}

   \authorrunning{H. Zhang \& W.-F. Yu }            
   \titlerunning{Effect of non-stationary accretion in 4U 1636-536}  
   \maketitle

%
%
\section{Introduction}           
\label{sect:intro}

Black hole transients usually exhibit different spectral states
during their outbursts, namely, the hard state, the soft state and
the state between the two, which is called the intermediate state or
the very high state \citep[see detailed definitions
in][]{rm2006,done2007}. Similar to black hole systems, according to
the source location in the X-ray color-color diagram, neutron star
low mass X-ray binaries also exhibit the hard state and the soft
state \citep{hasinger1989,yu2003,munoz-darias2014}. In current
popular theoretical models, X-ray spectral states are thought to be
determined primarily by the mass accretion rate, and there is a
critical threshold of the mass accretion rate -- below or above
which the accretion states, and therefore the spectral states, will
change \citep[e.g.][]{esin1997}. However, X-ray monitoring
observations show that the luminosity corresponding to the
hard-to-soft state transition during the rising phase of the
outbursts in LMXB transients is usually higher than the luminosity
corresponding to the soft-to-hard state transition
\citep{miyamoto1995,maccarone2003}, and this is also common in
neutron star soft X-ray transients \citep{bouchacourt1984,yu2003,
yu2007a}. This hysteresis effect indicates that the source can stay
in distinct spectral states at the same luminosity (or mass
accretion rate). Moreover, the hysteresis effect is mainly driven by
the large luminosity range in the hard-to-soft state transition. The
luminosity corresponding to the hard-to-soft transition has been
observed to vary by several times to one order of magnitude in
different outbursts of single sources \citep{yu2004,yu2007a}. These
observations indicate that there must be other parameters, other
than the mass accretion rate alone, which determine the source
spectral states and the state transitions. Some theories tried to
explain the hysteresis effect by considering the different amount of
Compton cooling or heating acting on the accretion disc corona
\citep{meyer-hofmeister2005} or by adjusting the viscosity parameter
$\alpha$\citep{qiao2009} or by considering the advection dominated
accretion flow with magnetically driven outflows \citep{cao2016}. In
these stationary accretion models, however,  the mass accretion rate
is still the primary tuning parameter which determines the accretion
or spectral state. Observations of most of the X-ray spectral state
transitions in Galactic X-ray binaries, especially those in LMXB
transients, do not support stationary accretion models for state
transitions.

A series of investigations of spectral state transitions based on
X-ray monitoring observations  indicate that non-stationary
accretion, especially those characterized by the large
rate-of-change of the mass accretion rate, plays an important role
in driving spectral transitions. \citet{yu2004},\citet{yu2007a} and
\citet{yu2007b} found a positive correlation between the hard X-ray
peak flux and soft X-ray peak flux for several low mass X-ray
binaries (Aql X-1, XTE J1550-564, 4U1705-44, and GX 339-4).
\citet{yu2009} and \citet{tang2011} performed a systematic study in
about 20 and 30 persistent and transient black hole and neutron star
X-ray binaries with the help of X-ray monitoring observations,
respectively, which confirmed the discovery in
\citet{yu2004,yu2007a} and \citet{yu2007b}, and found this
correlation holds over a luminosity range spanning by two orders of
magnitude. More importantly, they also found that the rate-of-change
of the X-ray luminosity around the spectral state transition is
positively correlated with both the luminosity corresponding to the
hard-to-soft state transition and the peak luminosity of the
outbursts. Motivated by these studies, \citet{zhang2015} found the
peak power of the episodic jet is positively correlated with both
the peak luminosity of the soft state (i.e., outburst peak
luminosity) and the rate-of-change of the typical X-ray luminosity
around the hard state during the rising phase of outbursts in
several black hole X-ray binaries in which good measurements of
source distances and black hole masses are available. The series of
investigations indicate that spectral state transitions, as well as
accretion or jet phenomena associated with the transitions, such as
episodic jets, are driven by non-stationary accretion instead of
stationary accretion. One of the key elements is the important role
of the rate-of-change in the mass accretion rate.

In this paper, we continue our investigation on spectral state
transitions in sources showing less dramatic change in the mass
accretion rates on the monitoring time scales. Our target 4U
1636-536, is a persistent neutron star LMXB often showing small
X-ray flares. It is classified as an atoll source by
\citet{hasinger1989}, suggesting that it usually evolves along the
spectral tracks between the island state (hard) and the banana state
(soft).  As seen with the {\it RXTE}/ASM, {\it Swift}/BAT and {\it
MAXI}/GSC in the last two decades, the X-ray intensity of this
source seen in X-ray monitoring observations shows flares up to a
factor of ten on the time scale of about 40 days \citep{belloni2007} in the past few
decades. It transits from the hard state to the soft state back and
forth during the flares. In this work we perform a systematic study
of fluxes corresponding to the hard-to-soft state transitions and
their relation to the peak fluxes of the corresponding soft states.


\section{Observations}
\label{sect:Obs}

The X-ray all-sky monitors, such as {\it Swift}/BAT, {\it RXTE}/ASM
and {\it MAXI}/GSC, have monitored 4U 1636-536 for decades on daily
time scales.  The monitoring light curves are publicly available.
This allows us to study the evolution of the X-ray spectral states
by studying the hardness ratio between the {\it Swift}/BAT and {\it
RXTE}/ASM intensity or between the {\it Swift}/BAT and {\it
MAXI}/GSC intensity. We take the {\it Swift}/BAT data obtained in
the period from 2005 February 12 (MJD 53413) to 2016 Jun 28 (MJD
57567), the {\it RXTE}/ASM data in the period from 2005 February 12
(MJD 53413) to 2010 April 18 (55304) and the {\it MAXI}/GSC data in
the period from 2010 April 19 (55305) to 2016 Jun 28 (MJD 57567).
Part of the {\it RXTE}/ASM and {\it Swift}/BAT has been reported in
\citet{yu2009} (from MJD 53413 to MJD 54504) and \citet{tang2011}
(from MJD 53413 to MJD 55304), however, we added more data from
the {\it MAXI}/GSC to further study the role of non-stationary accretion
in this source.

\section{Data reduction}
\label{sect:data}

All the X-ray intensity measurements have been converted into the
unit of Crab.  We set 1 Crab=0.23 count s$^{-1}$ cm$^{-2}$ for the
{\it Swift}/BAT and 1 Crab=75 count s$^{-1}$ for the {\it RXTE}/ASM.
For the case of {\it MAXI}/GSC, the 2-10 keV intensity was estimated
based on the combination of the
intensity of 4U 1636-536 in 2-4 keV and 4-10 keV,
by using the distance to be 6.0 $\pm$ 0.5 kpc \citep{galloway2008}
and making use of the measurements of the Crab X-ray luminosity in 2-10 keV
according to its energy spectrum with photon index of 2.07 and
normalization of 8.26 keV$^{-1}$ cm$^{-2}$ s$^{-1}$
\citep{kirsch2005}. The difference in the estimate of the fluxes in
Crab unit brought by using different energy bands, such as 2--12 keV
from {\it RXTE}/ASM and 2--10 keV from {\it MAXI}/GSC, is only about
3.7\%, which is smaller than the typical error of individual daily
measurements. Therefore the measurements with different monitors can
be consistently compared.

Though for the same count rate, different flux can be obtained
from different spectra and more and more evidences show that the
spectra of neutron star LMXBs are softer than the spectra of black
hole LMXBs \citep[e.g.][]{weng2015,wijnands2015}, the method we
converted the count rate to the flux by using the Crab Nebula spectrum is still 
reasonable. As \citet{yan2015} pointed out that the difference of
the flux estimated from the Crab Nebula spectrum to the flux
estimated from the typical soft and hard states are about 10\% and
25\% in the 2--12 keV, respectively. We also compared the flux
estimated from the Crab Nebula spectrum and the flux estimated by
fitting the quasi-simultaneous {\it RXTE}/PCA standard product of 4U
1636-536, and found the difference is less than 20\% when the source
reached the peak of the soft state. So the uncertainty of the flux
estimated from the Crab Nebula spectrum is less than about 25\%.

Figure 1 shows the light curves of 4U 1636-536 observed with these
all-sky monitors, which were binned to two-day time resolution to
get higher signal to noise ratio. The hardness ratio was defined as
the flux ratio between the {\it Swift}/BAT and {\it RXTE}/ASM or
{\it MAXI}/GSC. We followed the same method as described in
\citet{yu2009} and \citet{tang2011} to search for the state
transitions. According to the distribution of the hardness ratios of
neutron star LMXBs as shown in Figure 1 of \citet{yu2009}, the
hardness ratio thresholds for the hard state and the soft state of
4U 1636-536 are 1.0 and 0.2, respectively. The source is identified
in the hard state or the soft state when the hardness ratio is above
1.0 or below 0.2, respectively.  We identified the peak flux in the
hard state before the hard-to-soft state transition as the
transition flux based on the {\it Swift}/BAT light curve, and the
peak flux from the {\it RXTE}/ASM or {\it MAXI}/GSC light curves as
the peak flux of the following soft state. With these measurements,
we will study the relation between the transition flux of the
hard-to-soft state and the peak flux of the soft state in the source
in more details.

We notice that there are relatively good coverage of the {\it
RXTE}/PCA observations with about every two days from March 2005 to
the end of 2011, however, in order to try to explore the larger
parameter spaces of the transition flux of the hard-to-soft state
and the peak flux of the soft state, and to use a consistent method
as we have done in \citet{yu2009} and \citet{tang2011}, we use the
data which spannd more than eleven years long with the {\it
RXTE}/ASM, {\it MAXI}/GSC  and {\it Swift}/BAT. The results of our
selected data are very similar to the results of {\it RXTE}/PCA,
which are published in \citet{belloni2007} (See the {\it RXTE}/PCA
light curve in Figure 1 and hardness intensity diagram in Figure 2
of the paper) and \citet{munoz-darias2014} (See Fig.2 and 3 for the
hardness intensity diagram and the light curve, respectively.) The
difference of the thresholds of the hardness ratio between ours and
these two papers' are caused by the definition of the hardness ratio
with different energy band.

\begin{figure*}
\begin{center}
\centerline{\includegraphics[width=16cm,angle=0]{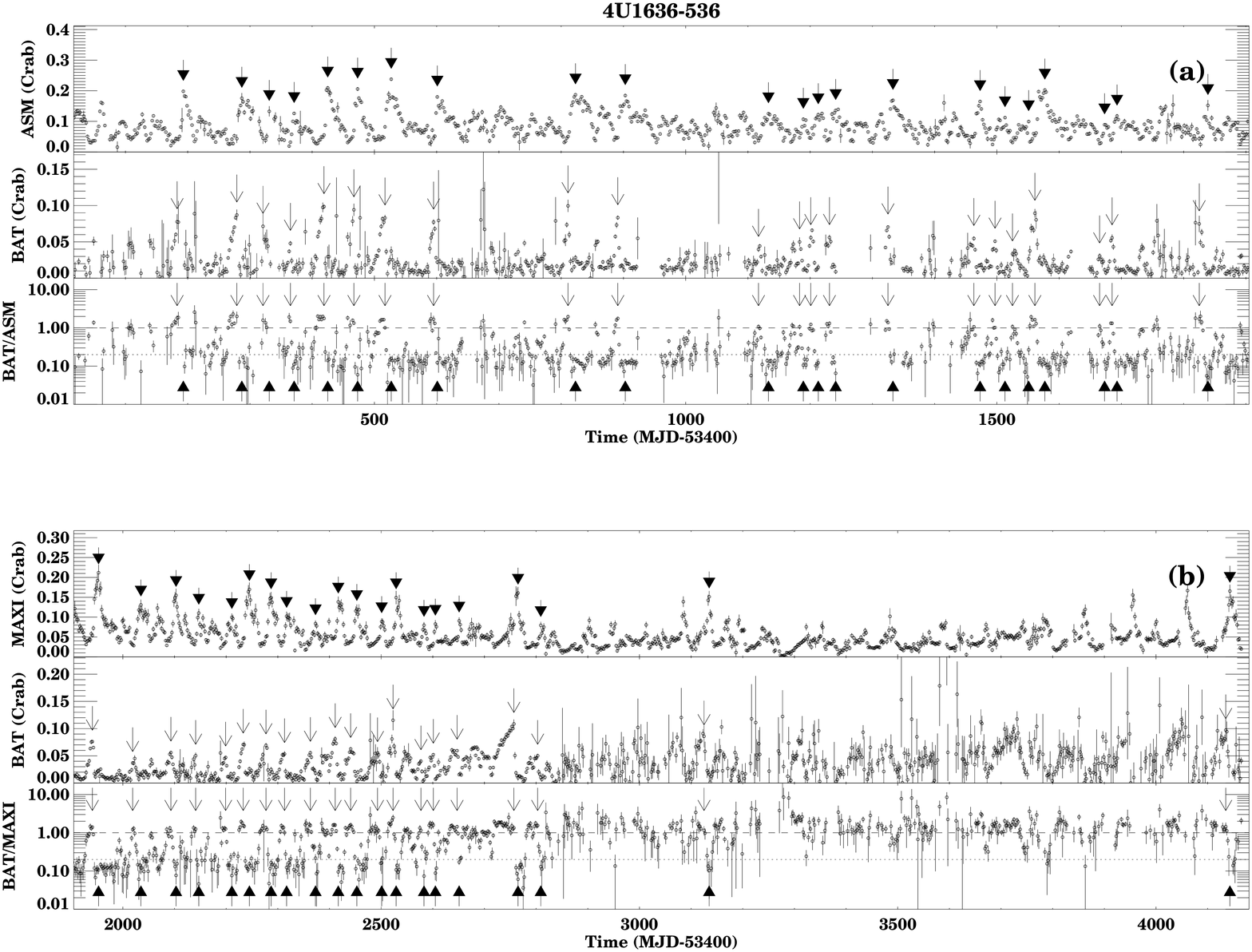}}
\caption{X-ray monitoring observations of 4U 1636-536 with the {\it
RXTE}/ASM in 2$-$12 keV, the {\it MAXI}/GSC in 2$-$10 keV and the
{\it Swift}/BAT in 15$-$50 keV. For each hard-to-soft state
transition, the starting time of the transition are marked in the
{\it Swift}/BAT light curve and the hardness ratio plot with thin
arrows, and the peak flux of the following soft state are marked  in
the {\it RXTE}/ASM or {\it MAXI}/GSC light curve and the
corresponding hardness ratio with thick arrows.}
\end{center}
\end{figure*}

\section{Results}

We have identified a total of 42 hard-to-soft state transitions in
4U 1636-536 in the period between 2005 February 12 and 2016 June 28.
There are a total of 22 hard-to-soft state transitions which were
identified with the the {\it Swift}/BAT and {\it RXTE}/ASM data as
are shown in Panel (a) of Fig. 1, and a total of 20 hard-to-soft
state transitions were identified with the {\it Swift}/BAT and {\it
MAXI}/GSC as are shown in Panel(b) of Fig.1.

\subsection{Correlation between the flux of the hard-to-soft state transition and the peak flux of the soft state}
Figure 2 shows the relation between the X-ray flux corresponding to
the hard-to-soft state transition in 15--50 keV and the peak flux of
the soft state in 2--12 keV. The data points shown in black and red
in Fig. 2 were obtained from the observations of {\it Swift}/BAT and
{\it RXTE}/ASM, and from the observations of {\it Swift}/BAT and
{\it MAXI}/GSC, respectively. The Spearman rank correlation
coefficient is 0.77, with the chance possibility of 2.1 $\times
10^{-9}$, indicating that there is a strong correlation between the
X-ray flux corresponding to the hard-to-soft state transition and
the peak X-ray flux of the following soft state. This is the direct
evidence that such a correlation holds in the single source. We fit
the data with a linear model of the form
log$F_{ps}$=$A$log$F_{tr}$+$B$, where $F_{ps}$ and $F_{tr}$
represent the peak flux of the soft state and the flux of the
hard-to-soft state transition, respectively. We obtained
$A$=1.07$\pm$0.06 and $B$=0.44$\pm$0.07. The index $A$ is almost the
same as that obtained by \citet{yu2009} for all the bright X-ray
binaries. The correlation is not affected by the uncertainties in
the estimates of the source distance and neutron star mass.
The X-ray fluxes were converted to the luminosities by setting the
source distance to 6.0$\pm$0.5 kpc\citep{galloway2008}. The
corresponding result is shown in Fig. 3. The large error of the luminosity are casued by
the uncertainty of the source distance. The errors of the transition
luminosities and the peak luminosities are not independent, so we do not fit the correlation,
but the correlation between them should be similar to the correlation between the transition flux
and the peak flux.

\begin{figure*}
\begin{center}
\centerline{\includegraphics[width=14cm,angle=0]{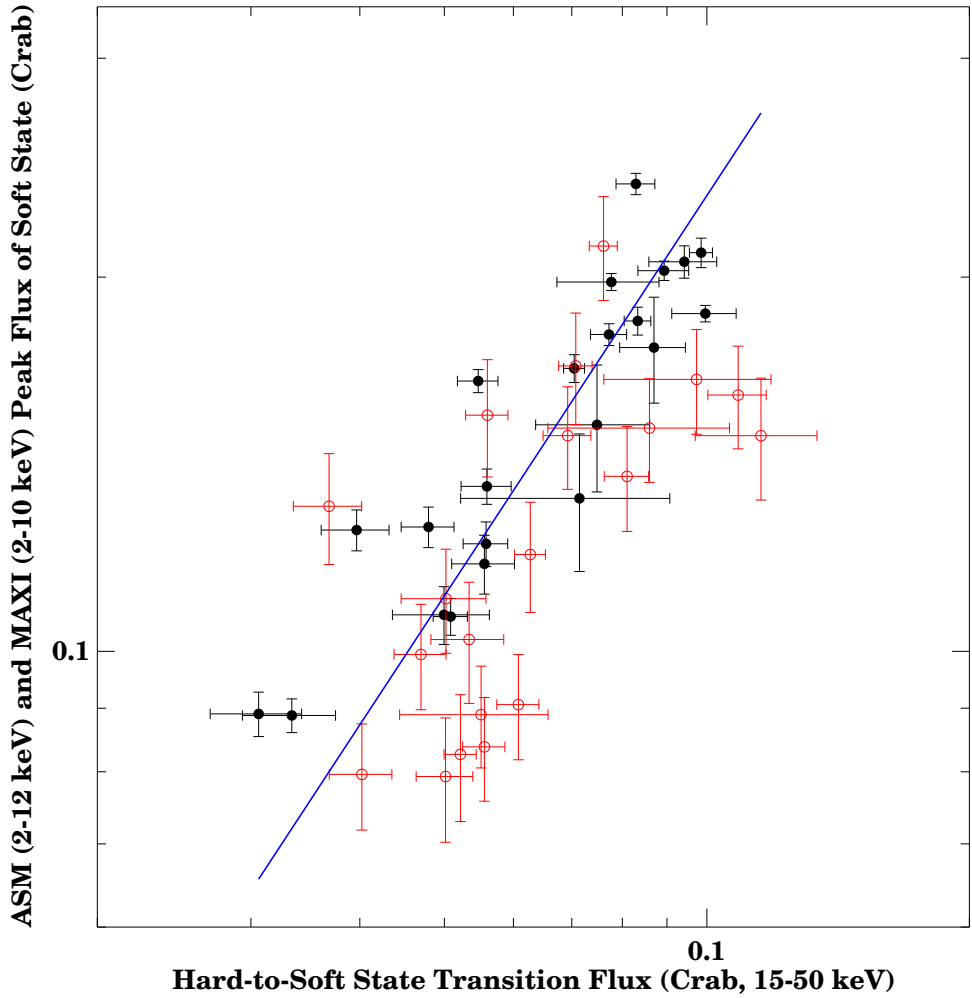}}
\caption{The correlation between the hard-to-soft state flux and the
peak flux of the soft state in 4U 1636-536. The filled black data
are from {\it RXTE}/ASM, and the red open circle are from {\it
MAXI}/GSC. The blue solid line is the linear fit result.}
\end{center}
\end{figure*}

\begin{figure*}
\begin{center}
\centerline{\includegraphics[width=14cm,angle=0]{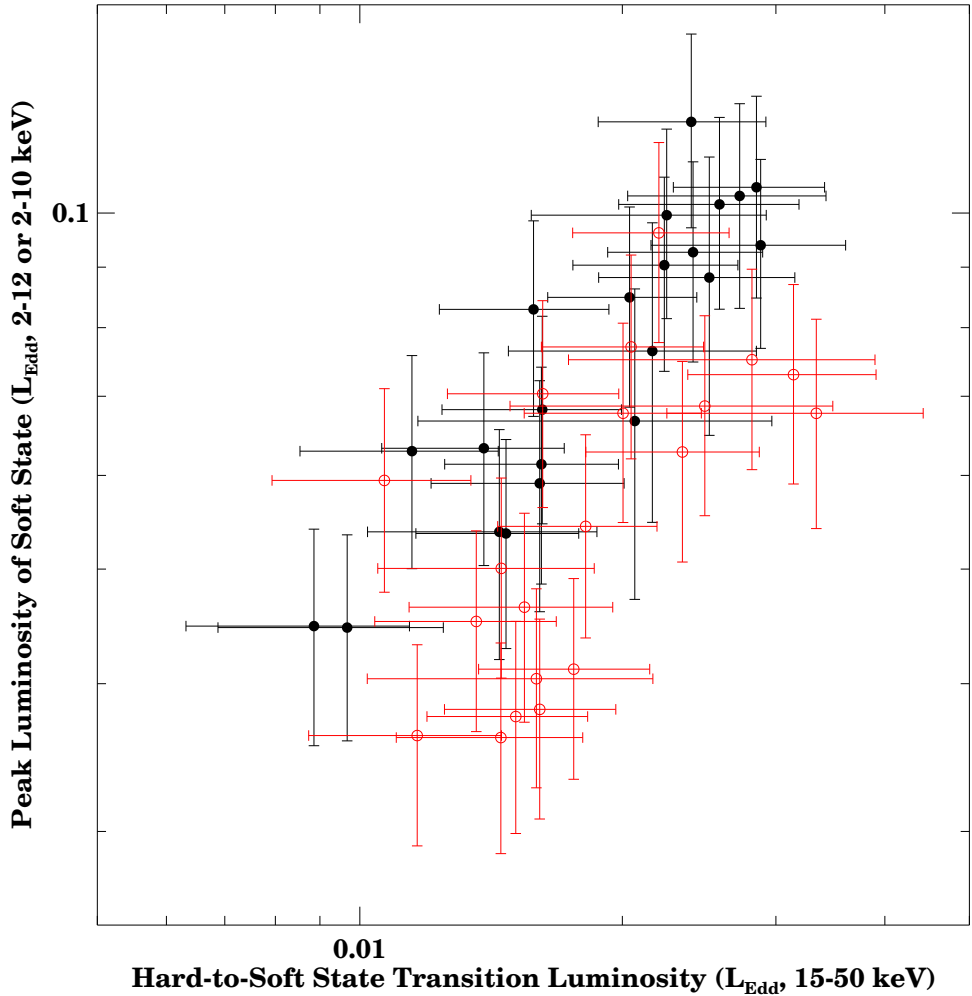}}
\caption{The same as Figure 2 with the unit in luminosity. }
\end{center}
\end{figure*}

In order to show the details of the spectral state transitions
during those X-ray flares, we show six typical X-ray flares with
good coverage of the X-ray observations in Fig. 4. Different symbols
represent different flares, the blue and red data points show the
light curves from the {\it Swift}/BAT and from the {\it RXTE}/ASM or
the {\it MAXI}/GSC, respectively. The data connected with solid or
dashed lines represent three brighter flares or three weaker flares,
respectively. In the plot, the positive correlation can be seen.
When the flux of the hard-to-soft state transition is higher in the
brighter flare than that in the weaker flare, the following peak
flux of the soft state is also higher in the corresponding flare, as
was also shown in Fig. 28 of \citet{yu2009}.

\begin{figure*}
\begin{center}
\centerline{\includegraphics[width=14cm,angle=0]{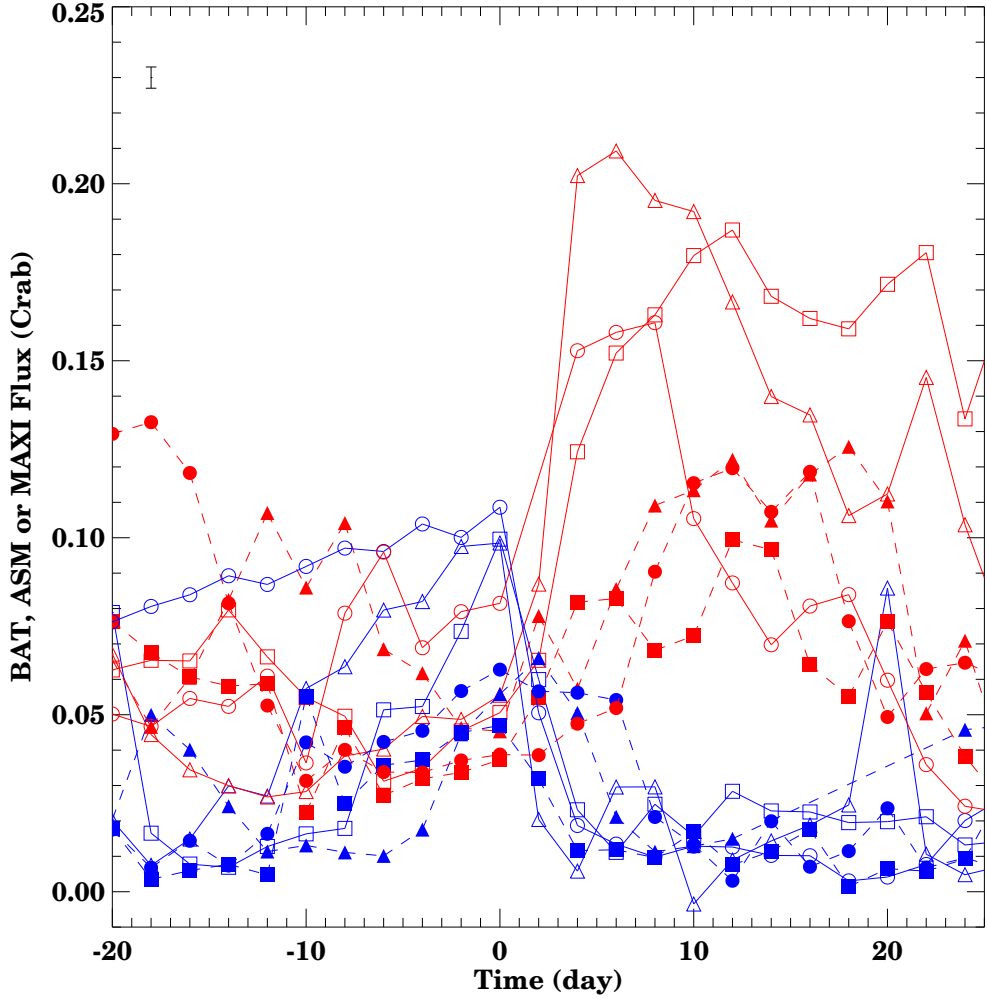}}
\caption{Examples of the light curves of six different flares. The
blue and red symbols represent the light curves of {\it Swift}/BAT
and {\it RXTE}/ASM or {\it MAXI}/GSC,
    respectively. The solid symbols connected with dashed lines and open symbols
    connected with solid lines represent typical weaker and brighter flares,
    respectively. Same symbols represent the same flares. The error bar shown 
	at the top left corner represents typical errors of the flux measurements. The time
    corresponding to the start of the hard-to-soft state transition
    is shifted to 0 for the light curves of {\it Swift}/BAT. All the light curves of
    {\it RXTE}/ASM or {\it MAXI}/GSC are shifted with the same time intervals
    corresponding to their {\it Swift}/BAT light curves. }
\end{center}
\end{figure*}

\section{Discussion}
\label{sect:discussion}

The luminosity corresponding to the hard-to-soft state transition
can vary by about a factor of 4 and the peak luminosity of the soft
state can vary by about a factor of 3 during the flares in 4U
1636-536. Though the outflows (wind or jet) will reduce the
actual mass accretion rate accreted to the central compact object
and the radiation efficiency may depends on the mass accretion rate
for different accretion flow \citep{narayan1998} , the variation of
the X-ray flux or luminosity can still roughly indicates the
variation of the mass accretion rate. Our results therefore indicate
that the hard-to-soft state transitions can occur at different
mass accretion rates in a large range in 4U 1636-536. This is 
contrary to theoretical models based
on stationary accretion which predicted the occurrence of the
hard-to-soft state transitions when the mass accretion rate exceeds
a critical mass accretion rate \citep[e.g.][]{esin1997}. 

The series of studies of the hard-to-soft spectral state transitions in black
hole and neutron star X-ray binaries demonstrate that we are in need
of non-stationary accretion models, which is characterized by the
consideration of rate-of-change of the mass accretion rate, for the
interpretation of spectral state transitions in X-ray binaries. 
Several studies showed that there was an obvious difference of the
X-ray spectra between the black hole and neutron star X-ray
binaries\citep[e.g.][]{vanderklis1994,shaposhnikov2009,farinelli2011,weng2015,
wijnands2015,burke2017}, but the non-stationary accretion plays an
important role in both of these two kinds of sources
\citep{yu2009,tang2011}. This is because the non-stationary
accretion is related to the process of the accretion instead of
the properties of the central compact object. There 
is no obvious difference of the hard-to-soft state transition luminosity 
in Eddingtion unit between the neutron star LMXB and the black hole LMXB. 
For example, the luminosity in Eddington unit of the hard-to-soft state 
transition in Aql X-1 is sometimes larger and sometimes lower than that in GX 339-4 
in different outbursts \citep[see Fig.24 in][]{yu2009}. \citet{lin2018} also concluded that the 
critical mass accretion rate of hard-to-soft state transition is not affected by the nature of
the surface of the compact stars. According to the framework of non-stationary accretion 
and its effect on spectral state transitions, the luminosity of the hard-to-soft transition is 
mainly determined by the scale of non-stationary accretion, revealing 
the effect of another parameter such as the rate-of-change of the mass 
accretion rate.

We found in 4U 1636-536, the average time interval between the
occurrence of the hard-to-soft state transition and the occurrence
of the flux peak of the soft state is about 9 days, but it can vary
in a range from 2 days to 16 days in single flares when the data are
binned on the time scales of two days.  This shortest time interval
might be challenging, since if it is consistent with the viscous
timescale at the inner edge of a standard thin disk approaching from
a large radius to the inner most radius, the initial radius should
be rather small. The challenge may be solved in future
non-stationary accretion models. Unfortunately, due to the data
quality, we could not measure the rate-of-change of the X-ray flux
in the hard state consistently among the flares, which led to large
uncertainties and strongly affected in individual cases. A
monitoring program with more sensitive X-ray observations in the
future would help achieve such a measurement.

\section{Conclusions}
\label{sect:conclusion}

In this work,  we report the discovery of a positive correlation
between the luminosity of the hard-to-soft state transition and the
peak luminosity of the following soft state in a single persistent
neutron star X-ray binary 4U 1636-536. This is consistent with the results of previous
studies on individual transient or quasi-persistent low mass X-ray
binaries \citep{yu2004,yu2007a,yu2007b}, and then all the bright
black hole and neutron star X-ray binary samples \citep{yu2009,tang2011}. This is the first
time that such a correlation is determined in a single persistent
neutron star X-ray binary with frequent small amplitude X-ray flares. Our result
indicates that there is no obvious difference of the effects of the
non-stationary accretion on spectral state transitions between soft
X-ray outbursts in LMXB transients and small X-ray flares in
persistent LMXBs except the flaring amplitude, which indicates that
our knowledge of the regimes of non-stationary accretion is
essential in understanding the properties of the accretion flow and
its accretion states.

\begin{acknowledgements}
We thank the {\it RXTE} and the {\it Swift} Guest Observer
Facilities at NASA Goddard Space Flight Center for providing the
{\it RXTE}/ASM products and the {\it Swift}/BAT transient monitoring
results, and the {\it MAXI} team for providing the {\it MAXI}/GSC
data.  This work was supported in part by the National Program on
Key Research and Development Project (Grant No. 2016YFA0400804) and
the National Natural Science Foundation of China (grant number
11103062, U1531130, and 11333005). WY would like to acknowledge the
support by the FAST Scholar fellowship.  The FAST fellowship is
supported by Special Funding for Advanced Users, budgeted and
administrated by Center for Astronomical Mega-Science, Chinese
Academy of Sciences (CAMS).  This work has made use of the data
obtained through the High Energy Astrophysics Science Archive
Research Center Online Service, provided by the NASA/Goddard Space
Flight Center.

\end{acknowledgements}

\label{lastpage}


\begin{thebibliography}{99}


\bibitem[Belloni et al.(2007)]{belloni2007} Belloni, T., Homan, J., Motta, S., Ratti, E., M{\'e}ndez, M., 2007, \mnras, 379, 247

\bibitem[Bouchacourt et al.(1984)]{bouchacourt1984}Bouchacourt, P., Chambon, G., Niel, M., Refloch, A., et
al., 1984, \apj, 285, L67

\bibitem[Burke et al.(2017)]{burke2017} Burke, M. J., Gilfanov, M., Sunyaev, R. 2017, \mnras, 466, 194

\bibitem[Cao(2016)]{cao2016} Cao, X. W., 2016, \apj, 817, 71

\bibitem[Done et al.(2007)]{done2007} Done, C., Gierli{\'n}ski, M.,
\& Kubota, A., 2007, A\&AR, 15, 1

\bibitem[Esin et al.(1997)]{esin1997} Esin, A. A., McClintock, J. E., Narayan, R., 1997, \apj, 489, 865

\bibitem[Farinelli \& Titarchuk(2011)]{farinelli2011}Farinelli, R., \& Titarchuk, L. 2011, A\&A, 525, 102

\bibitem[Galloway et al.(2008)]{galloway2008} Galloway, D. K., Muno, M. P., Hartman, J. M., Psaltis, D., \& Charkrabarty, D., 2008, \apjs, 179, 360

\bibitem[Hasinger \& van der Klis(1989)]{hasinger1989} Hasinger, G., \& van der Klis, M., 1989, A\&A, 225, 79

\bibitem[Kirsch et al.(2005)]{kirsch2005} Kirsch, M. G. et al., 2005, Proc.SPIE, 5898, 22

\bibitem[Lin \& Yu (2018)]{lin2018} Lin, J. \& Yu, W. F. 2018, \mnras, 474, 1922
%
\bibitem[Maccarone \& Coppi (2003)]{maccarone2003} Maccarone, T. J., \& Coppi, P. S., 2003, \mnras, 338,
189

\bibitem[Meyer-Hofmeister et al.(2005)]{meyer-hofmeister2005} Meyer-Hofmeister, E., Liu, B. F., Meyer, F., 2005, A\&A, 432, 181

\bibitem[Miyamoto et al.(1995)]{miyamoto1995} Miyamoto, S., Kitamoto, S., Hayashida, K., Egoshi,
W., 1995, \apj, 442, L13

\bibitem[Mu{\~n}oz-Darias et al.(2014)]{munoz-darias2014} Mu{\~n}oz-Darias, T.; Fender, R. P.; Motta, S. E.; Belloni, T. M., 2014, \mnras, 443, 3270

\bibitem[Narayan et al.(1998)]{narayan1998}Narayan, R., Mahadevan, R., \& Quataert, E. 1998, in Theory of Black Hole
    Accretion Disks, ed. M. A. Abramowicz, G. Bjornsson, \& J. E. Pringle
(Cambridge: Cambridge Univ. Press), 148

\bibitem[Qiao \& Liu(2009)]{qiao2009} Qiao, E. L., \& Liu, B. F., 2009, PASJ, 61, 403

\bibitem[Remilliad \& McClintock(2006)]{rm2006} Remillard, R. A. \&
McClintock, J. E., 2006, ARA\&A, 44, 49

\bibitem[Shaposhnikov \& Titarchuk(2009)]{shaposhnikov2009} Shaposhnikov, N., \& Titarchuk, L. 2009, \apj, 699, 453

\bibitem[Tang et al.(2011)]{tang2011} Tang, J., Yu, W. F., \& Yan, Z., 2011, RAA, 11, 434

\bibitem[van der Klis (1994)]{vanderklis1994} van der Klis, M. 1994, \apjs, 92, 511

\bibitem[Weng et al. (2015)]{weng2015} Weng, S. S., Zhang, S. N., Yi, S. X., Rong, Y., \& Gao, X. D. 2015, \mnras, 450, 2915

\bibitem[Wijnands et al. (2015)]{wijnands2015} Wijnands, R., Degennar, N., Armas Padilla, M., et al. 2015, \mnras, 454, 1371

\bibitem[Yan \& Yu (2015)]{yan2015} Yan, Z., \& Yu, W. 2015, \apj, 805, 87

\bibitem[Yu et al.(2003)]{yu2003}Yu, W. F., Klein-Wolt, M., Fender, R., van der Klis, M.,
2003, \apj, 589, L33

\bibitem[Yu et al.(2004)]{yu2004} Yu, W., van der Klis, M., Fender, R., 2004, \apj, 611 , L121

\bibitem[Yu et al.(2007)]{yu2007b} Yu, W., Lamb, F. K., Fender, R. P., van der Klis, M., 2007, \apj, 663, 1309

\bibitem[Yu \& Dolence(2007)]{yu2007a} Yu, W., Dolence, J., 2007, \apj, 667, 1043

\bibitem[Yu \& Yan(2009)]{yu2009} Yu, W., \& Yan, Z., 2009, \apj, 701, 1940

\bibitem[Zhang \& Yu(2015)]{zhang2015} Zhang, H., \& Yu, W. F., 2015, \mnras, 451, 1740




\end{thebibliography}
\end{document}